\documentclass[12pt,preprint]{aastex}

\shorttitle{Dynamics of Multiple Stellar Formation Events}
\shortauthors{Downing \& Sills}

\begin{document}

\title{The Dynamical Implications of Multiple Stellar Formation Events
  in Galactic Globular Clusters}

\author{Jonathan M.B. Downing \altaffilmark{1} and Alison Sills}
\affil{Department of Physics and Astronomy, ABB-241, McMaster University,
  1280 Main Street West, Hamilton, Ontario, L8S 4M1, Canada}
\altaffiltext{1}{Current Address: Astronomisches Rechen-Institut, Zentrum
  f\"ur Astronomie Universit\"at
  Heidelberg, M\"onchhofstra\ss e 12-14, D-69120, Heidelberg, Germany}
\email{downin@ari.uni-heidelberg.de, asills@mcmaster.ca}

\begin{abstract}

Various galactic globular clusters display abundance anomalies that affect the
morphology of their colour-magnitude diagrams.  In this paper we consider the
possibility of helium enhancement in the anomalous horizontal branch of NGC
2808.  We examine the dynamics of a self-enrichment scenario in which an
initial generation of stars with a top-heavy initial mass function enriches
the interstellar medium with helium via the low-velocity ejecta of its
asymptotic giant branch stars.  This enriched medium then produces a second
generation of stars which are themselves helium-enriched.  We use a direct
N-body approach to perform five simulations and conclude that such
two-generation clusters are both possible and would not differ significantly
from their single-generation counterparts on the basis of dynamics.  We find,
however, that the stellar populations of such clusters would differ from
single-generation clusters with a standard initial mass function and in
particular would be enhanced in white dwarf stars.  We conclude, at least from
the standpoint of dynamics, that two-generation globular clusters are
feasible.

\end{abstract}


\section{Introduction}

Globular clusters are gravitationally bound collections of stars,
typically including between $10^{3}$ and $10^{7}$ members, which normally
occur within the halos of galaxies.  Galactic globular clusters are
understood to be old objects, up to 12 Gyr in age \citep{GCages}, and are thus
primordial components of the galaxy.  In the simple picture, all stars within
a globular cluster are thought to have formed at the same time and out of the
same medium, thus sharing the same age and chemical composition.  In
particular all stars in a given globular cluster have the same value of
$[Fe/H]$ and are said to be mono-metallic.  The only differentiation between
stars in this scenario is their spectrum of masses determined by an initial
mass function (IMF).  There is thought to be a universal IMF for all galactic
globular clusters but there is some debate as to its exact form
(\cite{KroupaIMFrev} and references therein).

Although this description works well for the most part, several globular
clusters display population characteristics that are inconsistent with this
model.  In many clusters all stars have the same $[Fe/H]$ but there are
star-to-star abundances variations in other elements. In particular many
globular clusters stars show strong star-to star O-Na and Mg-Al
anticorrelations \citep{GCabund}.  Some clusters also display peculiar
morphologies in the horizontal branch of the Colour-Magnitude Diagram (CMD).
In some cases different clusters with the same $[Fe/H]$ and with otherwise
identical colour-magnitude diagrams will have differing horizontal branch
morphologies (e.g. the case of M3 and M13 (\cite{M3M13}, \cite{M3M13FeH}).
This is called the ``second parameter problem'' since such clusters seem to
require a second parameter (other than metalicity) in order to explain the
variations between them. In some cases, the presence of very extended
horizontal branch blue tails is linked to dynamical or abundance effects.  One
possible source of these anomalies is a self-enrichment scenario where
chemicals produced by some cluster stars are incorporated into other cluster
stars which then display unusual characteristics such as enhancements of light
and heavy elements (particularly products of hot-hydrogen burning and the CNO
cycle) and unusual colours (frequently bluer) with respect to field stars (see
the review by \cite{GCabund} for further details).  We propose to explore the
dynamics of a particular variation of the self-enrichment scenario in which
excess helium is produced by asymptotic giant branch (AGB) stars and which may
generate the anomalous horizontal branch of the globular cluster NGC 2808.

NGC 2808 is a southern galactic globular cluster first observed in detail in
the late 1960s \citep{first2808}.  It is quite massive at $1.6 \times 10^{6}
M_{\odot}$, has a moderately high velocity dispersion $\sigma_{o} = 13.4 km/s$
\citep{PandM} and is relatively metal rich with $[Fe/H] = -1.09$
\citep{Harriscat}.  As early as 1974 a peculiar morphology (both bimodal and
extended) was discovered in the horizontal branch (HB) of the CMD of NGC 2808
\citep{Harris74}.  The horizontal branch of NGC 2808 consists of both a small
horizontal clump (hearafter called the red horizontal branch or RHB) and a
separate, extended tail reaching vertically from 16th to 22nd magnitude
(hereafter called the blue horizontal branch or BHB).  The BHB is also
multi-modal with three distinct groups (called from top to bottom extended
blue tails (EBTs) 1, 2 and 3).  This is very different from canonical HB
morphology (flat and monomodal) and in order to explain it a special scenario
is required.

One way to explain the anomalous morphology of the HB of NGC 2808 is to assume
that the BHB stars are helium enriched while the RHB stars contain the
cosmological helium abundance of $Y \sim 0.24$ \citep{DandC2002}.  The
enrichment cannot be attributed to helium variations across the primordial
cloud since the necessary levels of enrichment are extreme ($Y \sim 0.32 -
0.4$ is needed in EBT 3 \citep{DandC2004}, \citep{2808HeMs}) and the helium
must arise from stellar processes through self-enrichment. It has been argued
\citep{Gnedin2002} that the more massive globular clusters (particularly
$\omega$-Centari) may have sufficient self-gravity to retain the low-velocity
ejecta from AGB stars (and in the case of the most massive even some of the
ejecta from Type II supernova).  In the case of clusters within $\sim$10 kpc
(such as $\omega$-Centari) \cite{Gnedin2002} argue that interaction with the
galactic disc would strip such ejecta from clusters before it can experience
star formation.  NGC 2808, however, is 11.1 kpc from the galactic center
\citep{Harriscat} and may avoid such stripping long enough to undergo a second
generation of star formation \citep{DandC2004}.  This leads
to a scenario in which helium self-enrichment from AGB stars produces the
horizontal branch morphology of NGC 2808.  In particular \cite{DandC2004}
propose a model where NGC 2808 has experienced two stellar formation events.
The first event features a top-heavy mass function that enhances the number of
$3 - 5 M_{\odot}$ stars.  These evolve to the asymptotic giant branch (AGB)
phase in $\leq 200$ Myr and enrich the interstellar medium (ISM) with helium
(and other elements) through stellar winds.  A second generation of stars then
forms from the helium-enriched ISM.  An age difference of $\leq 200$ Myr would
not be observable because it is a small fraction of the lifetime of a low mass
star, but the helium difference in the populations would be.  The low-mass
stars in the first generation would form the RHB of NGC 2808 and the enriched
second generation would form the BHB.  The multi-modality of the BHB (EBTs 1,
2 and 3) can be explained by varying levels of helium enrichment produced
either by several minor stellar formation events producing different levels of
helium enrichment or possibly by differential helium enrichment across the
cluster.  Modeling more than two generations or trying to account for
spatially dependent helium enhancement is beyond the scope of our simulations
and in this paper we do not investigate the substructure in the BHB.  It is
possible that helium enrichment is the source of other anomalous CMD
morphologies.  Notably the double main sequence of the massive cluster
$\omega$ Centauri can be explained if the blue main sequence stars are helium
enriched \citep{PiottoCen}\footnote{$\omega$ Centauri has many other 
  peculiarities and may be the core of a stripped dwarf
  galaxy.  It cannot be considered typical of galactic globular
  clusters}.  Indeed \cite{2808HeMs} claim helium enrichment in some 20\% of
main sequence stars for NGC 2808 (although this case is less clear cut) but we
have not tried to model this effect.

Based on stellar evolution models \cite{DandC2004} claim that, given certain
assumptions, sufficient helium can be produced by AGB stars in the
two-generation scenario to explain the BHB of NGC 2808.  If enhancements in
other elements produced by AGB stars are ignored and the top-heavy mass
function is accepted, the scenario may be plausible on the basis of stellar
chemistry.  The massive first generation may, however, have effects on both
the dynamics and population of the cluster. Specifically the high mass of the
initial generation and the effect of adding a second generation to a
dynamically evolved object could lead to dynamical instabilities or affect the
final spatial distribution of a star cluster.  In addition the first
generation will leave many intermediate-mass remnants, primarily white dwarfs,
which should still be observable after 10-12 Gyr (e.g. \cite{Hansen02}).  We
use a direct N-body approach to explore the dynamical evolution of a globular
cluster in the D'Antona and Caloi two-generation formation scenario.

\section{Method}

To perform our simulations we use Starlab, a direct N-body evolution code.
Starlab uses a fourth-order Hermite predictor-corrector scheme
\citep{McMillan86} and contains a full suite of stellar and binary evolution
algorithms based on the work of \cite{Eggelton89}.  Starlab is described in
some detail in \cite{Starlab01} and is freely available with some
documentation from http://www.ids.ias.edu/\~{}starlab/.  We also use a GRAPE-6A
N-body gravitational accelerator \citep{Baby} to improve the speed and size of
our simulations.

\subsection{Top-Heavy IMF}

\begin{figure}[!hbp]
\centering
\plotone{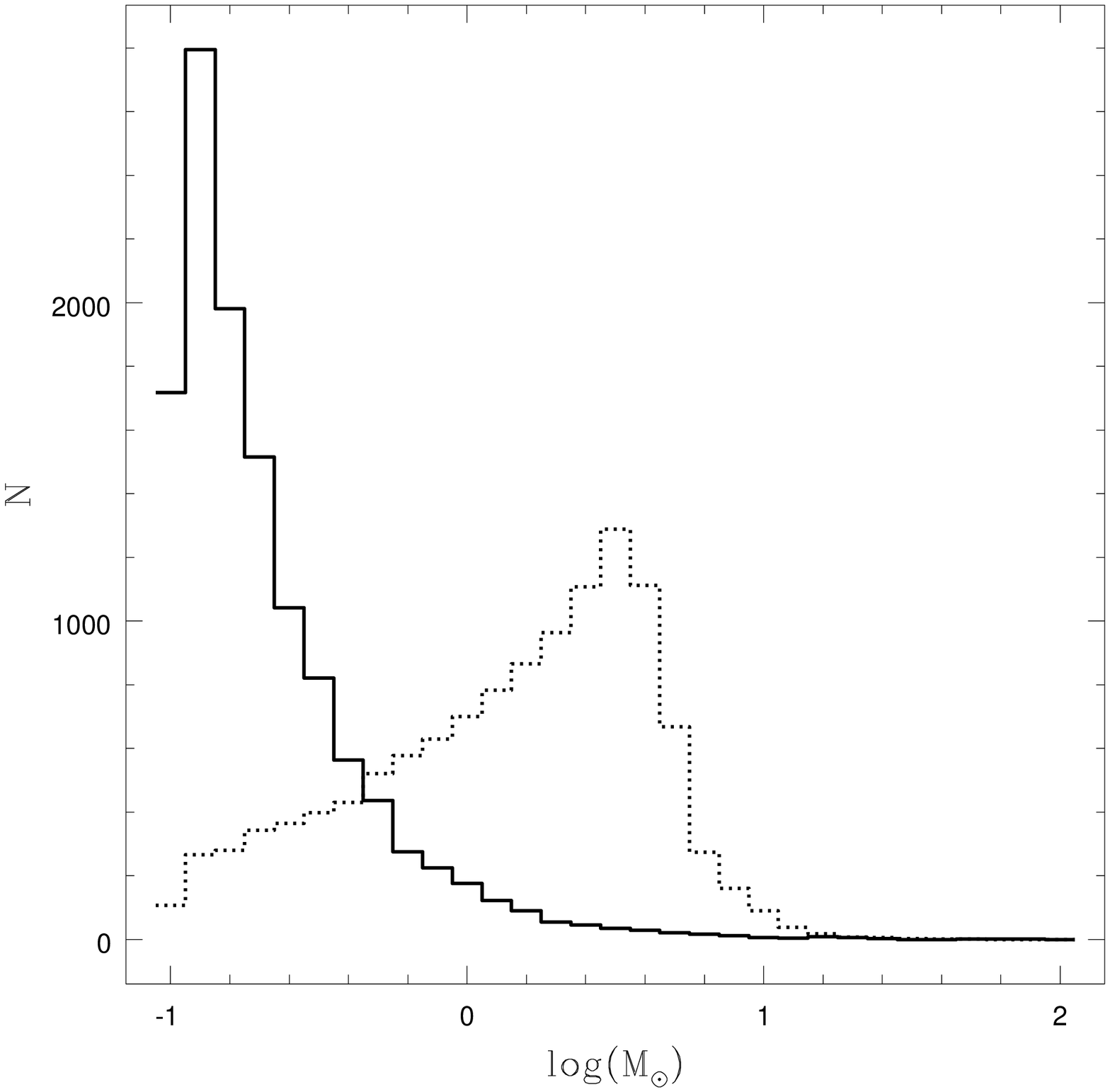}
\caption[D'Antona and Caloi vs. Salpeter IMF]{A D'Antona and Caloi IMF
  (dotted line) prepared by our new Starlab algorithm compared with a
  Salpeter IMF prepared using one of the standard Starlab algorithms
  (solid line). \label{fig:DCSalmf}}
\end{figure}

The D'Antona and Caloi top-heavy IMF given in \cite{DandC2004} is a broken
power-law which takes the form:
\begin{displaymath}
\frac{dN}{dM} = \left\{ \begin{array}{ll}
c_{1}M^{-(1+\alpha)} & \textrm{if $M \leq M_{B}$} \\
c_{2}M^{-(1+\beta)} & \textrm{if $M > M_{B}$}
\end{array} \right.
\end{displaymath}
and is defined by the exponents $\alpha$ and $\beta$ and a transition mass
$M_{B}$ ($c_{1}$ and $c_{2}$ are normalization constants).  The explicit
implementation of this IMF in Starlab is discussed in Appendix A.  A D'Antona
and Caloi IMF created with this implementation where $\alpha = -0.5$, $\beta =
3.0$ and $M_B = 3.8$ is compared with a standard Salpeter IMF in Figure
~\ref{fig:DCSalmf}.  These parameters correspond to the IMF from
\cite{DandC2004} with the lowest overall mass still capable of providing the
requisite helium enhancement.  This is ostensibly the least extreme scenario
and, since discrepancies between simulations of single-generation clusters and
more extreme examples of the D'Antona and Caloi scenario might not preclude
agreement in less extreme cases, we use these parameters in all of our
simulations.

\subsection{Details of the First Generation}

In modeling the first generation there are three compontents to consider, one
stellar and two gaseous.  The stellar component is simply the stars of the
first generation and can be dealt with by standard N-body modeling.  The first
gas component consists of the remnant gas from the first generation of star
formation.  The second gas component is composed of the ejecta from the AGB
stars of the first generation which produces the helium enhancement for the
second generation.

There is little information available on the configuration of remnant gas
(first component) in young globular clusters and for simplicity we model the
remnant ISM left over from the first generation of star formation as an
analytic Plummer potential, $\phi(r) = -\frac{GM}{\sqrt{r^{2}+a^{2}}}$.  This
potential is not static since the remnant ISM is stripped from the young
cluster by the radiation from young O stars, early supernovae and possibly due
to interactions with the galactic disk.  The exact rate of gas stripping from
young clusters is unknown but it on the order of 10s of Myrs
(\cite{Goodwin1997}, \cite{GasML2006}).  It is clear, however, that this
mass-loss affects early cluster evolution since the potential in which the
cluster lives becomes more shallow over time and the cluster can expand and in
the extreme case be completely disrupted.

The stellar ejecta forming the second component originates primarily
from supernovae and the stellar winds of AGB stars.  According to
\cite{Gnedin2002} the characteristic terminal velocity for AGB winds
is $\sim$ 15 km s$^{-1}$ whereas the escape velocity for a cluster of
the mass of NGC 2808 is on the order of 60 km s$^{-1}$ in the core and
40 km s$^{-1}$ at the half mass radius.  It is plausible that a large
fraction of the mass lost by AGB stars will in fact be retained as gas
by the cluster.  In our models we assume that all of the mass lost due
to stellar evolution is retained, that is, we do not treat ejecta from
different progenitor stars separetly.  This assumption is not
necessarily realistic since the high-velocity supernovae ejecta may be
able to escape from the potential well of the young cluster and some
of the AGB ejecta may be stripped by radiation in the same way as the
remnant ISM.  The majority of stars in the first generation are below
$10 M_{\odot}$ and will not undergo supernovae events in the first 200
Myr of the cluster's life.  Therefore the AGB ejecta should be the
dominant component and including the supernovae ejecta is not a
concern.  The stripping fraction for the AGB ejecta is also not clear
and treating it would vastly expand our parameter space without
providing any particular physical insight.  Thus the choice of a 100\%
retention fraction for the ejecta is a reasonable first approximation
and is also in the spirit of \cite{DandC2004}.  We model the retained
ejecta as another single, global Plummer potential with a mass
increasing at the rate mass is lost from the stars due to stellar
evolution.  To be more realistic, we should add gas to the simulation
locally where it is lost by the stars.  Since at $15 km s^{-1}$ the
gas crossing time at the initial viral radius is $\sim 3$ Myr (about
two simulation time units) and the particle crossing time is $\sim
8.5$ Myr (about four simulation time units) we expect gas produced
locally to be quickly distributed throughout the cluster.  Thus the
assumption of a global profile for the second gas component seems
reasonable.

\begin{figure}[!hbp]
\centering
\plottwo{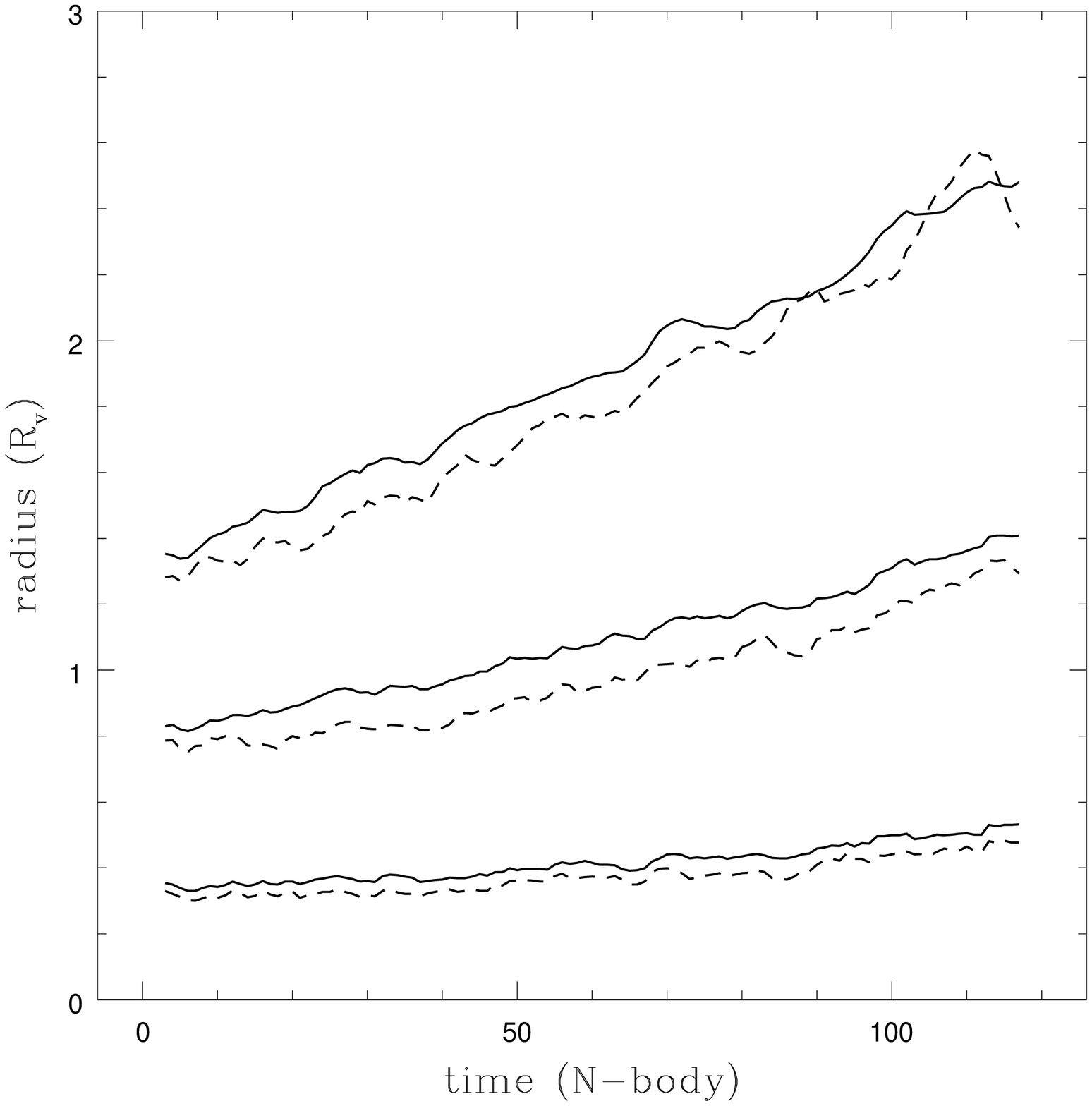}{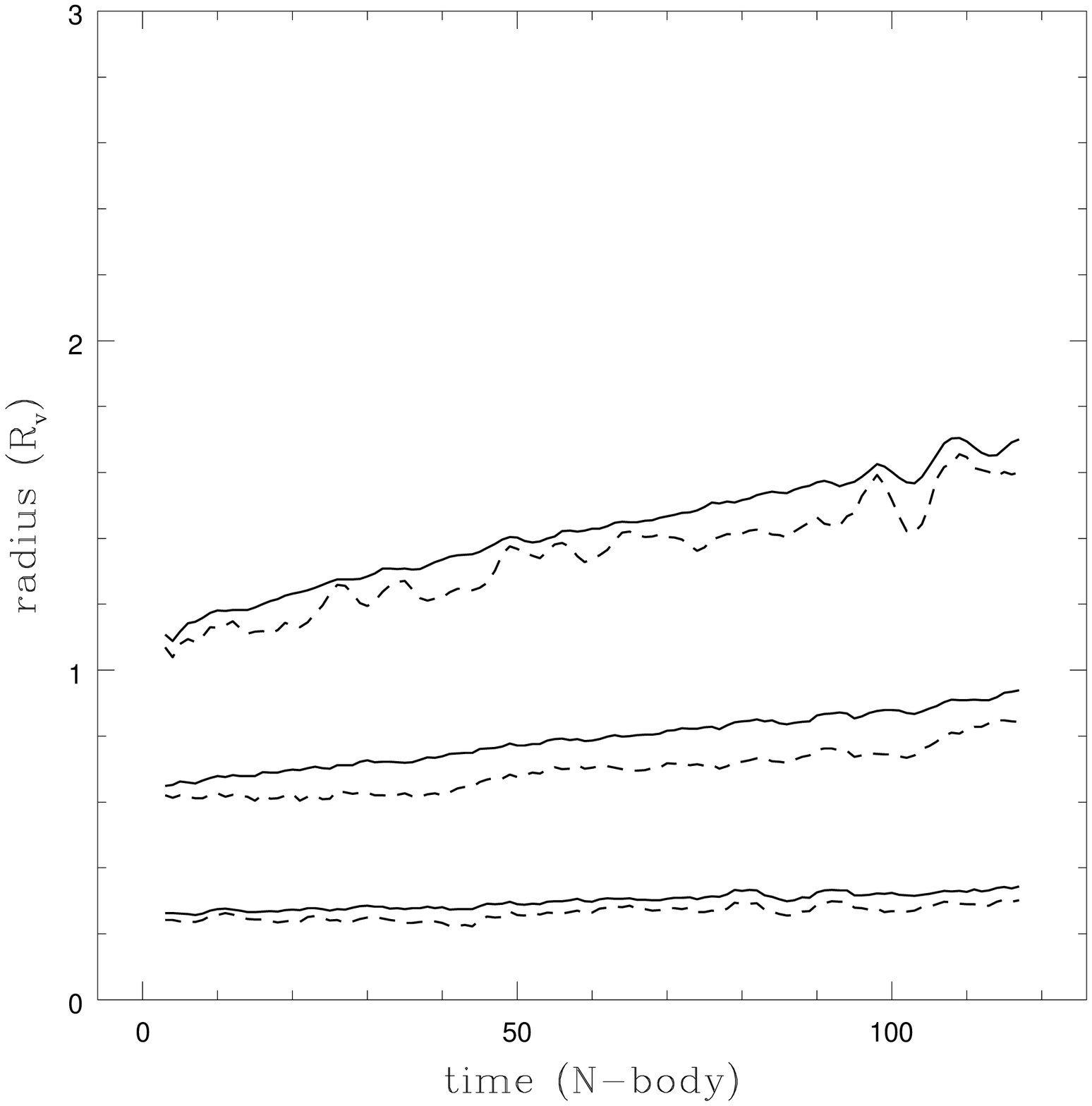}
\caption[The effect of a Plummer potential]{On the left are shown the
  Lagrangian radii for a cluster with a top-heavy IMF evolving for 200
  Myr.  On the right are the Lagrangian radii for the same cluster evolving in
  a Plummer potential with a static mass equal to the mass lost to stellar
  evolution over the 200 Myr life of the cluster.  From bottom are the 10\%,
  50\% and 75\% Lagrangian radii.  Solid indicates the entire system, dotted
  indicates the most massive 10\% of stars. \label{fig:pfeffect}}
\end{figure}

Since Starlab does not contain prescriptions for time-evolving potentials and
to incorporate such prescriptions linked both to radiation pressure and
stellar evolution would be time-consuming we make the following
approximations: (1) We assume that after 200 Myr all of the remnant ISM has
been removed from the cluster.   (2) We assume that the remnant ISM is
stripped from the cluster at the same rate at which ejecta from the first
generation stars is added to the cluster.  (3) We assume that the initial mass
of the first component is the same as the final mass of the second component.
This leads to a model where the first-generation is embedded in a $static$
Plummer potential with a total mass equal to the mass lost by stellar
evolution in the first generation.  Assumption (1) is almost certainly secure
since young clusters appear to be stripped of their remnant star forming
material on timescales of a few 10s of Megayears but this means assumption
(2) is not necessarily accurate since the remnant ISM will be ejected more
rapidly than it is replaced by the ejecta from the first generation (added to
the cluster over a timescale of 200 Myr).  We argue that the two interacting
potentials will act in three phases.  Initially the remnant ISM potential will
provide a deep well which will cause the cluster to contract.  Once the
remnant ISM has been removed and before the ejecta potential becomes large the
cluster will be able to expand normally due to mass-loss.  Finally the ejecta
potential will dominate and strongly mitigate the effect of mass-loss and
again slow the expansion.  The global static potential will have the correct
behaviour at the beginning and end and will simply slightly deepen the
potential in the middle phase.  In both cases the dynamical effect should
simply be to suppress the expansion of the cluster due to stellar-evolution
mass-loss.  Assumption (3) is simply convenient and, since the exact star
formation rate in young clusters is unknown, any variation of this paramater
would not provide any new physical insight.  To confirm the static potential
has no additional dynamical effects, we compare the evolution of one of our
first generations with and without a potential in Figure~\ref{fig:pfeffect}.

As expected the only qualitative difference between the two plots in
Figure~\ref{fig:pfeffect} is that with the addition of the Plummer potential
both the massive and overall Lagrangian radii increase more slowly and lead to
a smaller, more compact final cluster.  This effect is seen in both the
massive and overall Lagrangian radii and is a good indication that mass
segregation has not been suppressed by the addition of the external field.
Since mass-segregation is a two-body relaxation effect \citep{MassSeg} this
shows that two-body relaxation also has not been affected by the addition of
the Plummer potential.

Obviously a pair of time-evolving potentials linked to the stellar
physics of the code would be preferable but the difference would be in
details rather than in overall effect.  Any galactic globular cluster
will have experienced $\sim10$ Gyr of dynamical evolution since the
potential formation of a second generation and thus be a fully
dynamically relaxed object.  Since dynamically relaxed N-body systems
are insensitive to their initial conditions \citep{Binney&Tremaine} we
would not expect fine details of the gas potential in the first
generation to be significant for current observations. We note that
our predictions are not applicable to two-generation clusters in the
process of forming their second generation or even those less than a
relaxation time old.  For such objects, however, it is not clear that
global potentials would be sufficient to accurately resolve the
details of evolution and such models may have to wait until hybrid
SPH-N-body codes are available.

\subsection{The Second Generation}

In order to match current observations we assume the second generation forms
with a more standard IMF and for simplicity we choose a Salpeter IMF
\citep{Salpeter55} .  In order to implement the second generation we must find
some way add stars to the output file from the simulation of the first
generation (in Starlab such a file is know as a snapshot).  The easiest way to
implement this in Starlab is to create a separate snapshot using the
parameters characterizing the second generation and then combining this
snapshot with the output snapshot from the first generation using the routine
{\tt merge\_snaps}.  We choose the total mass of the second-generation to
match the mass of the Plummer potential and hence the mass lost in our
first-generation.  This implicitly implies 100\% star formation efficiency.
This assumption is almost certainly unrealistic (e.g. \cite{LandL}) but it is
both in keeping with the spirit of the scenario proposed by \cite{DandC2004}
and it allows us to conserve mass when adding the second generation (all of
the Plummer potential mass is replaced by stellar mass).  Since the
actual star formation efficiency during galactic globular cluster formation is
not known, considering different star formation rates would again increase
our parameter space without providing any particular physical insight.  We use
the {\tt merge\_snaps} routine to combine the first and second generation
snapshots into a single snapshot and manually remove the Plummer potential
from the first generation.  This procedure represents the conversion of a
continuous gas potential into a generation of stars with the same total mass
in one timestep ($\sim1.5$ Myr).  Once the second generation has been added
and the Plummer potential removed, the combined cluster is evolved for $\sim
12$ Gyr which corresponds to 15-16 initial half-mass relaxation times
($T_{RH}$) \citep{Spitzer87}.  In order to produce simulations lasting more
than 2-4 initial half-mass relaxation times it is necessary to introduce a
stripping radius.  This is because newly-formed neutron stars are ejected from
the cluster with a high velocity due to their initial kick.  Once far enough
away, the code seems to treat pairs of such neutron stars as binaries and the
three-body interaction between these and the rest of the cluster eventually
reduces the timestep to zero and halts the simulation.  Removing stars at a
large radius from the cluster eliminates this problem.

\section{Initial Conditions}

We perform five simulations:  a single generation model with a
Salpeter IMF (SP), a single generation model with a D'Antona and Caloi
top-heavy IMF (DC), and three two-generation models with a D'Antona and Caloi
IMF for the first generation and a Salpeter IMF for the second generation (Ca,
Cb, and Cc).  In all cases we follow \cite{Hurley05} and use Plummer models
for our initial conditions.  In the two-generation models we allow the first
generation to evolve for $\sim200$ Myr before adding the second generation
according to a Salpeter IMF.  In all cases the initial virial radius is the
length unit and is set to 5.0 pc.  The mass unit for each simulation is the
total mass of the simulated cluster.  The energy and time units are defined by
starting all simulations in virial equilibrium.  Parameters for these
simulations are given in Table~\ref{tab:InPar}.  The number of particles used
in these simulations is a comprimise between achiving an accurate number of
particles and our hardware limitations.  \cite{F&H95} have explored the
behavior of clusters with different mass funcitons and find that there is
little difference between simulations carried out with 8192 and 16384
particles.  Practically, we find that $\sim 12000$ particles is the maximum
number for which we can achive a reasonable wall-clock time and since we are
in the middle of the 8000-16000 particle range, we do not expect our results
to scale to strongly with particle number.  All clusters are
simulated in isolation in order to keep our parameter space small and to
eliminate effects that are not due to the addition of the second generation.
We must, however, introduce a stripping radius for the reasons discussed
above.  We find that by stripping at 100 initial viral radii we eliminate the
problematic fast-moving neutron stars while removing few other stars from the
simulations.

\section{Results}

We consider results for the dynamical and final stellar populations of our
simulations separetly.

\subsection{Dynamics}

We present the initial half-mass relaxation times and some of the
final physical data for our systems in Table~\ref{tab:FinPar}.  Each cluster
has lost a significant portion of its initial mass.  Few particles are removed
from the simulations and this mass loss is due to stellar evolution rather
than loss of stars from the cluster.  This mass-loss is particularly
pronounced in DC with its extreme initial enhancement of high-mass stars and
is also a major factor in the two-generation clusters with their top-heavy
first-generation IMFs.  We find the initial half-mass relaxation times of the
two-generation clusters to be somewhat smaller than that of the single
generation Salpeter IMF while the half-mass relaxation time of the
single-generation cluster with the D'Antona and Caloi IMF is much shorter than
any of the other cases.  Since each cluster has the same initial virial radius
and thus the same initial physical concentration, the differences are
due to the initial total mass increasing from SP through the two-generation
clusters to DC.  It is worth noting that the combined clusters undergo a
sufficient number of relaxation times to allow the possibility of core collapse
by the end of our simulations.

\begin{figure}[!hbp]
\centering
\plottwo{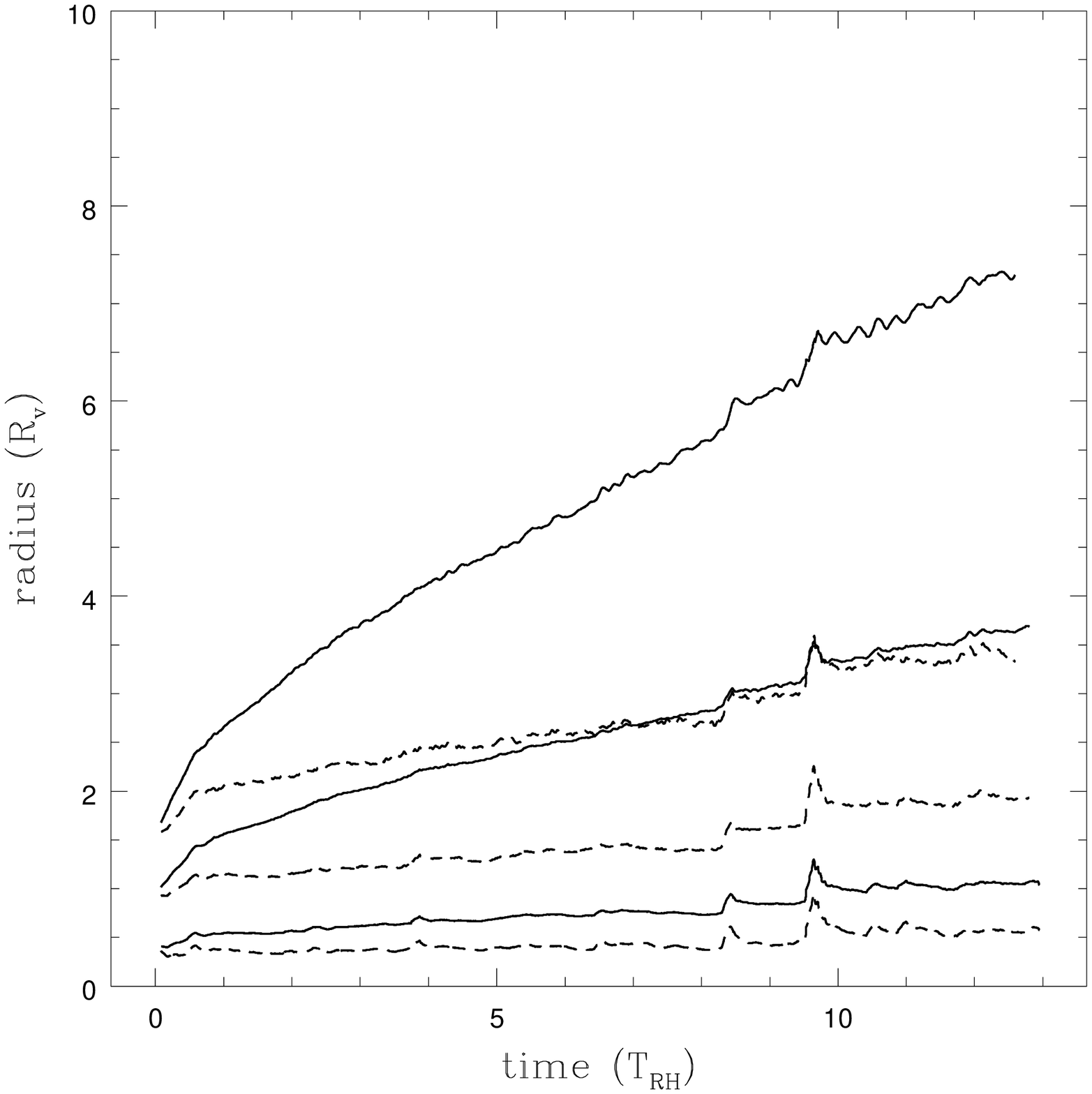}{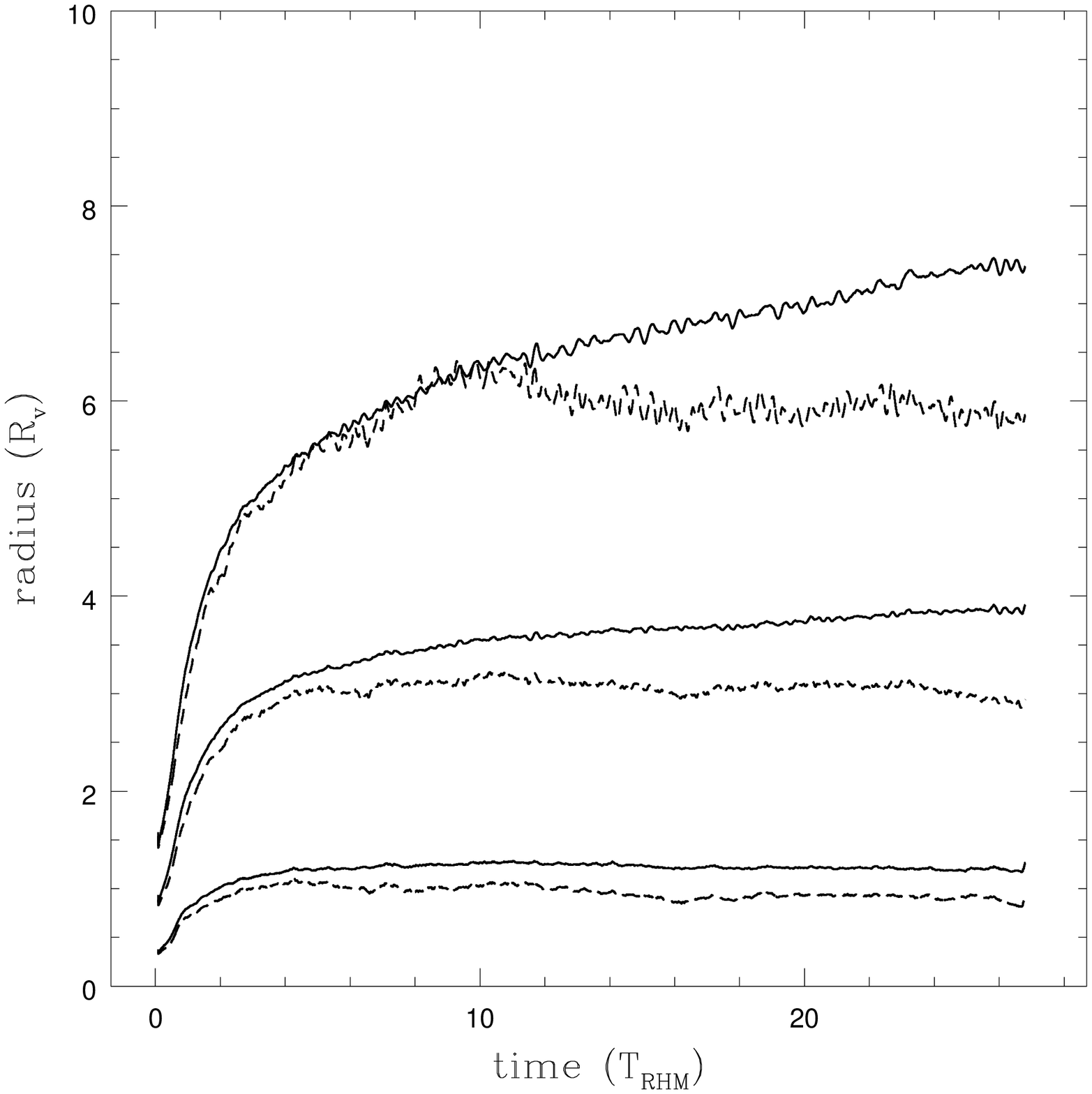}
\epsscale{0.45}
\plotone{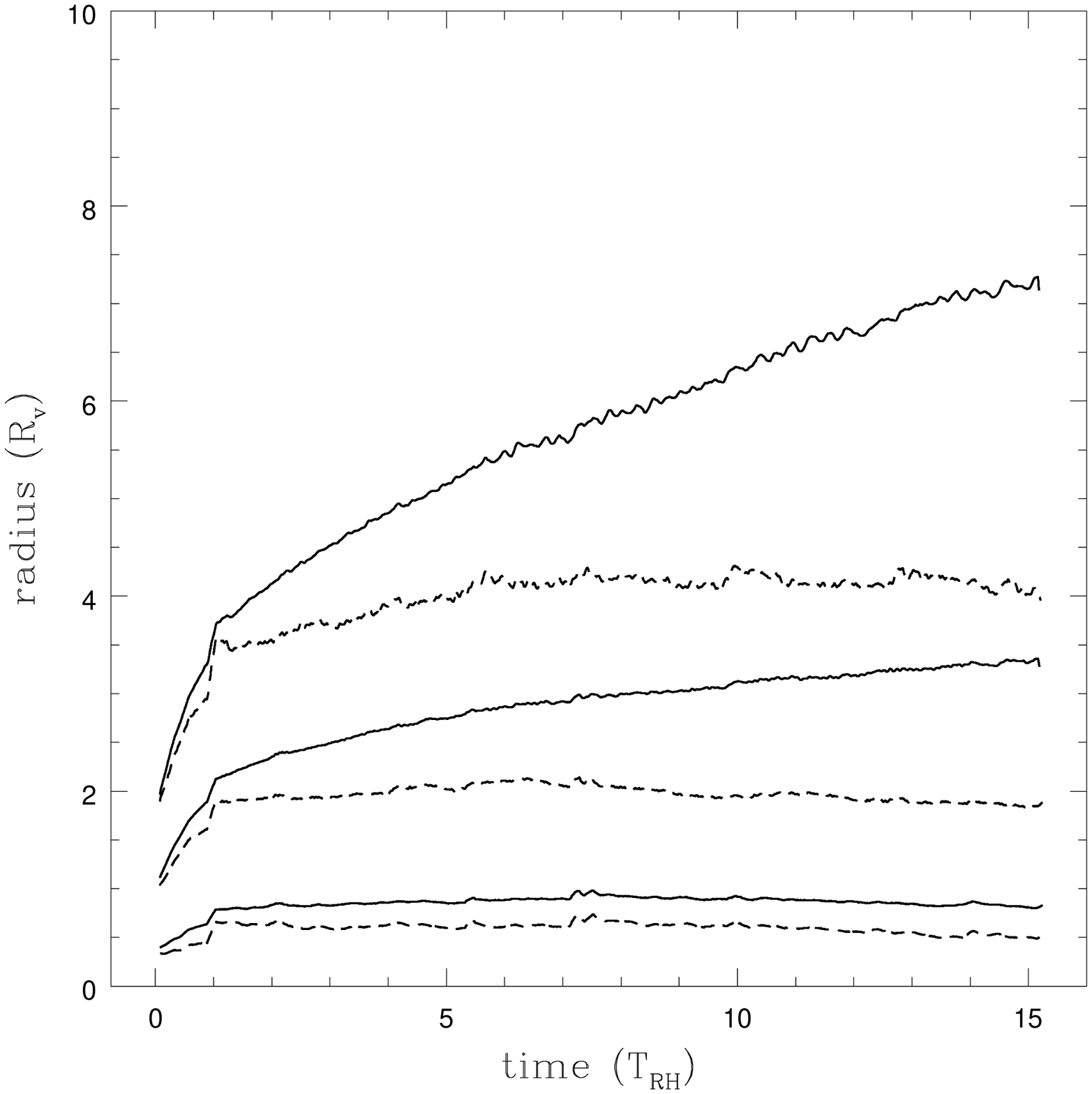}
\epsscale{1.0}
\caption[Lagrangian Radii]{The Lagrangian radii for a Salpeter IMF (top left),
  a D'Antona and Caloi IMF (top right) and a D'Antona and Caloi IMF with a
  second generation added according to a Salpeter IMF (bottom).  From
  bottom to top in each graph are the 10\%, 50\% and 75\% Lagrangian radii.
  Solid is for all stars, dotted for the most massive 10\% of stars.  The
  spike just before 10 $T_{RH}$ in SP is created by the ejection of a massive
  binary from the cluster core. \label{fig:Lagr}}
\end{figure}

In Figure~\ref{fig:Lagr} we present a comparison of the
Lagrangian radii for the runs SP, DC and Ca (Cb and Cc are similar to Ca).
The evolution of the Lagrangian radii for SP is slightly different from Ca and
DC.  In DC and Ca the Lagrangian radii grow quickly at first and then flatten
significantly, with the 75\% radius still growing slowly and the 50\% and 10\%
flattening (and even contracting in the case of the 10\% and 50\% massive
radii of Ca).  By contrast the early growth period is much less evident in SP
and all radii continue to grow slowly (with the 10\% radius flattening at the
end).  The difference is due to the high-mass stars in DC and the first
generation of Ca.  These high-mass stars evolve quickly and consequently lose
a great deal of mass over a short timescale (the first generation of Ca loses
$\sim30\%$ of its mass in 200 Myr).  This mass-loss leads to a rapid reduction
in cluster potential, causing rapid cluster expansion and a consequent
increase in the Lagrangian radii.  Once the high-mass stars evolve away,
mass-loss becomes dominated by low-mass stars which evolve much more slowly
leading to a much reduced mass-loss rate and a subsequent flattening of the
Lagrangian radii.  By contrast SP is always dominated by low-mass stars and
experiences slow, constant mass-loss throughout its life, consequently
displaying slow, constant expansion.

We compare the Lagrangian radii of the most massive 10\% of stars (cutoff
masses given in Table~\ref{tab:FinPar}) to the Lagrangian radii for the entire
system in Figure~\ref{fig:Lagr}.  We note that there is a wide separation
between the massive and overall Lagrangian radii in SP and Ca while for much
of the life of DC the massive and overall Lagrangain radii are similar.  That
the massive Lagrangian radii are smaller indicates the high-mass stars are
more centerally concentrated than the low-mass ones.  This is a manifestation
of mass segregation which is a natural consequence of equipartition of energy
in phase space and is expected in all N-body systems with a range in initial
masses \citep{Binney&Tremaine}.  The smaller spread between the overall and
massive population in DC seems to indicate that mass segregation is less
advanced in this simulation than in SP and Ca.  This is somewhat unexpected
since for an N-body system the timescale for mass segregation is on the order
of the two-body relaxation time \citep{Binney&Tremaine}.  Since DC has the
shortest half-mass relaxation time of our simulations we would expect mass
segregation would be the most advanced in this model.  We postulate that the
larger spread and higher overall mass in the IMF of DC
(Figure~\ref{fig:DCSalmf}) both renders the overall statistics of this
simulation less sensitive the the effects of a few very high-mass stars in the
central regions and produces a more smooth transition between the low-mass
halo and the high-mass core.  Both of these processes could bring the overall
and massive Lagrangian radii closer together without actually indicating a
reduction in mass-segregation.  It is worth noting that mass segregation
indeed does become more pronounced in DC after 5-10 $T_{RH}$ after the most
massive stars have evolved away and the mass function has become more peaked.
SP and Ca both exhibit extensive mass segregation based on the Lagrangian
radii with SP exhibiting a slightly larger discrepancy between the massive
and overall radii than Ca.  Again Ca has a shorter two-body relaxation
timescale than SP and we invoke the same idea as in DC to explain this
effect. Thus, since both SP and Ca exhibit a similar size at each overall
Lagrangian radius, we do not expect two-generation clusters to be reliably
distinguishable from their single-generation counterparts by observations of
overall size or mass segregation.

\begin{figure}[!hbp]
\centering
\plotone{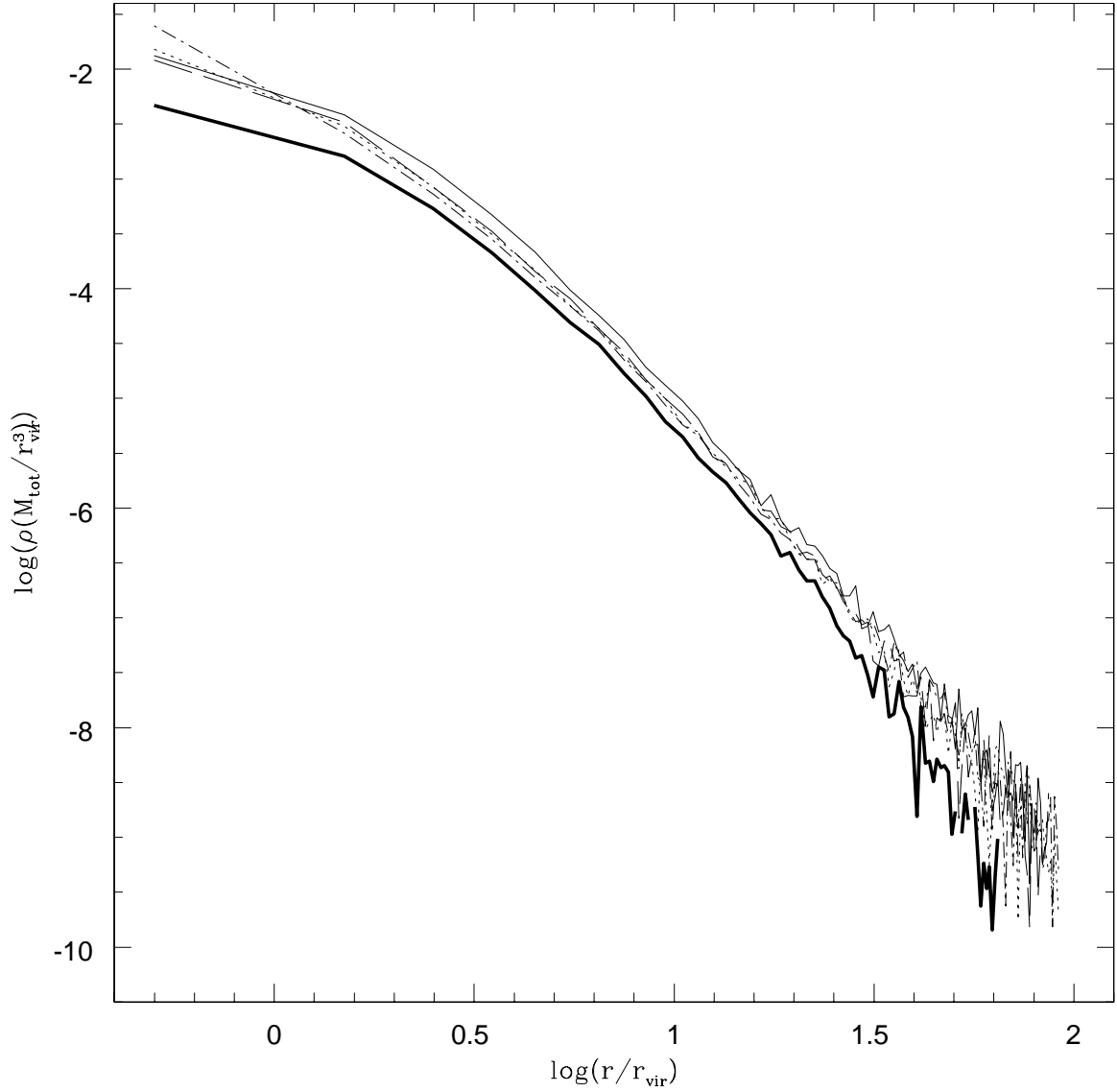}
\caption[Density Profiles]{The radial density profiles of the simulated
  clusters.  Thin solid is SP, thick solid is DC, dotted is Ca, broken is Cb
  and dot-dashed is Cc.  Profiles are taken to the stripping radius (100
  initial $R_{vir}$). \label{fig:DenProf}}
\end{figure}

In Figure~\ref{fig:DenProf} we consider the final density profiles of all
simulated clusters.  Note that the 75\% Lagrangian radius is at
$\log{r/r_{vir}} \approx 1$ and the profile is noisy outside this region due to
low-number statistics.  The density profiles are remarkably similar with all
clearly exhibiting an exponential halo and a flatter core.  Ca and Cb
are almost indistinguishable from SP except at the very center.  DC is slightly
less dense than the others but has essentially the same profile.  The only
simulation which displays a significant deviation is Cc with a steeper core
profile than the other simulations.  Cc differs from the other two-generation
clusters in other ways and will be discussed in more detail in \S~\ref{Scc}.

\begin{figure}[!hbp]
\centering
\plotone{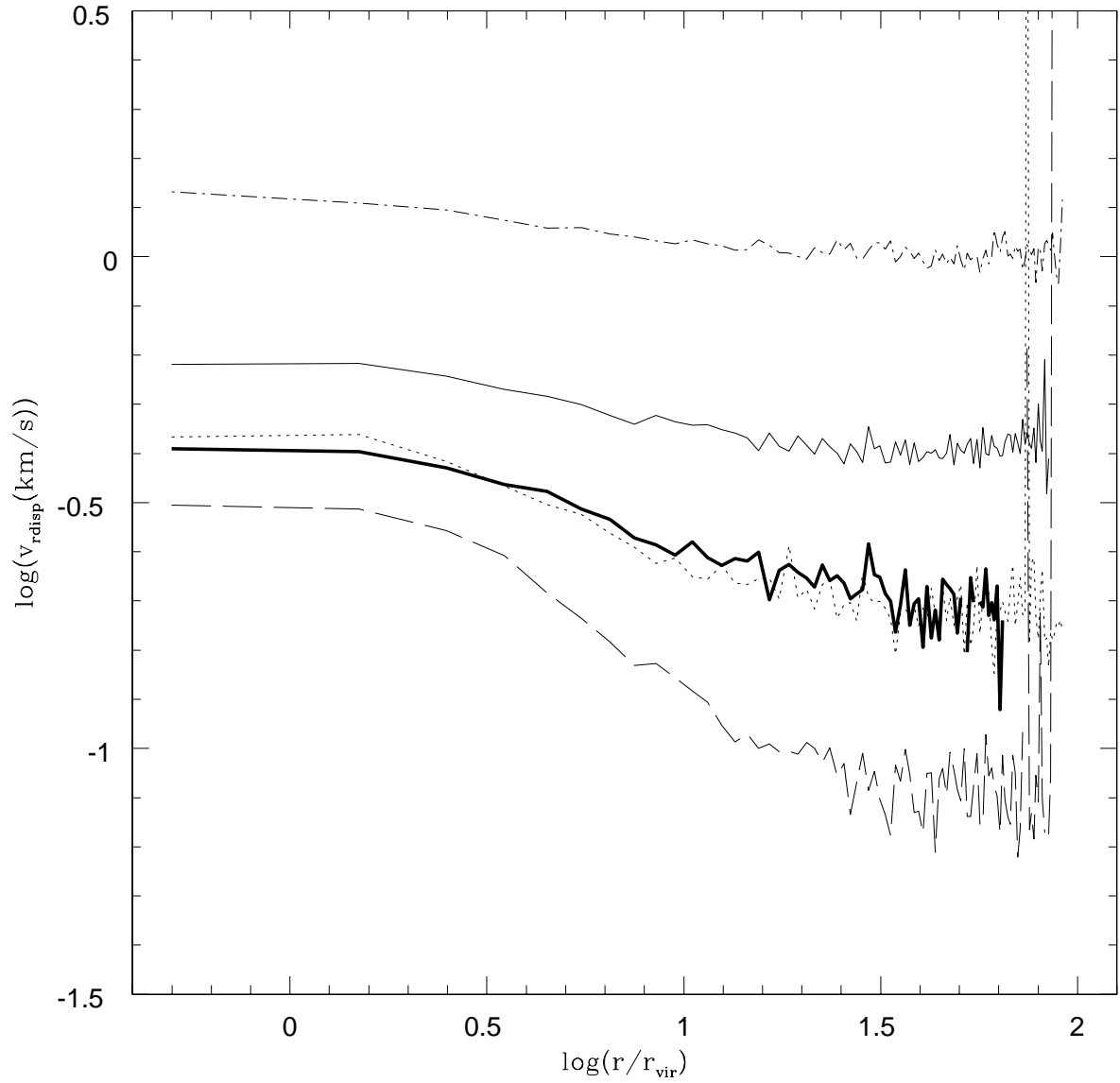}
\caption[Radial Velocity Dispersion]{Profiles of the radial velocity
  dispersion for the simulated clusters.  Thin solid is SP, thick solid is DC,
  dotted is Ca, broken is Cb and dot-dashed is Cc.  Profiles are taken to the
  stripping radius (100 initial $R_{vir}$).  \label{fig:RvdProf}}
\end{figure}

In Figure~\ref{fig:RvdProf} we consider the radial velocity dispersion profile
for all simulated clusters.  Again the 75\% Lagrangian radius is at
$\log{r/r_{vir}} \approx 1$ and statistics beyond this are point unreliable
due to low numbers.  We note that the cores of Ca and Cb are slightly more
distinct than SP and DC and that all profiles are vertically shifted compared
to each other.  These variations are as large between the two-generation
clusters themselves as between the two-generation and single-generation
clusters and they are merely simulation-to-simulation variations.
The peaks in Ca and Cb are caused by high-velocity single stars and are a
product of low-number statistics.  The largest outlier is again Cc and we now
turn to a discussion of this simulation.

\subsubsection{Simulation Cc}
\label{Scc}

\begin{figure}[!hbp]
\centering
\plotone{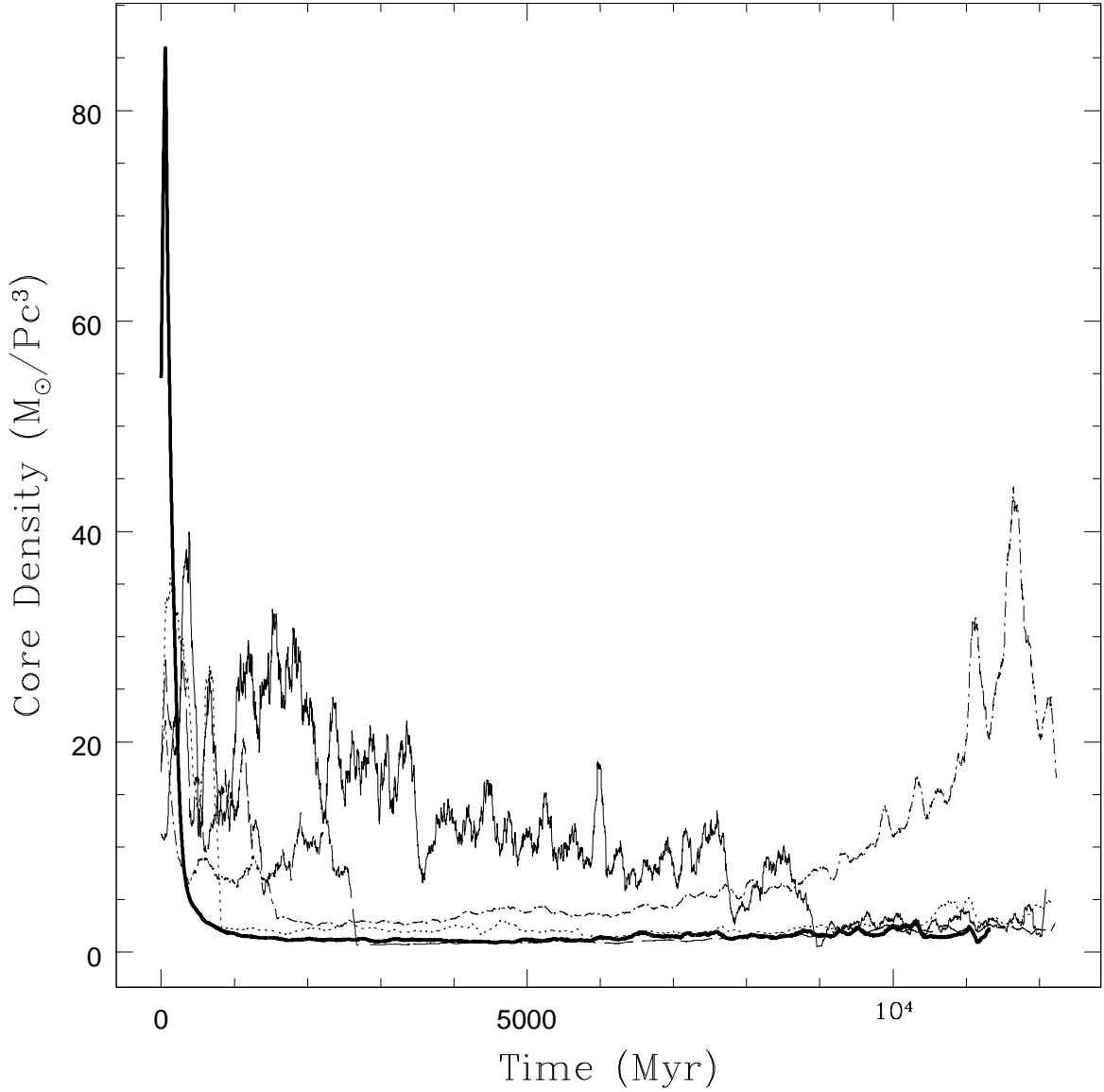}
\caption[Core Densities]{Core densities for the simulated clusters as a
  function of time.  Thin solid is SP, thick solid is DC, dotted is Ca, broken
  is Cb and dot-dashed is Cc.  Time is physical so that all simulations can
  appear on the same graph. \label{fig:CoreDen}}
\end{figure}

We consider core densities as a fucntion of time for all simulations in
Figure~\ref{fig:CoreDen}.  Unlike the other simulations, Cc exhibits a clear
peak in core density at approximately 15 $T_{RH}$ ($\approx 11.5$ Gyr),
corresponding nicely to the steeper density profile in the core region in
Figure~\ref{fig:DenProf}.  We tentatively propose that Cc has experienced a
core-collapse event at 15 $T_{RH}$.  If correct, this is an interesting result
and tells us that the dynamics of two-generation clusters are similar to
those of their single-generation counterparts in yet another way.  That only
one simulation would experience core collapse at a specific time is not
surprising since the onset of core collapse is associated with individual
escapers and hard binaries, the specifics of which vary between
simulations.  We caution, however, that with only 12000 particles such
a peak could be caused by a few massive stars in the core region rather than a
full core-collapse event.  We also note that the kinetic energy of escapers in
a core-collapsed cluster should be systematically higher than in a
non-core-collapsed cluster due to binary scatting from the core.   This is not
the case for Cc (Table~\ref{tab:FinPar}).  Further simulations with more
particles are needed to fully pin down the core-collapse behaviour of
two-generation clusters. 

\subsection{Stellar Population}

\begin{figure}[!hbp]
\centering
\plottwo{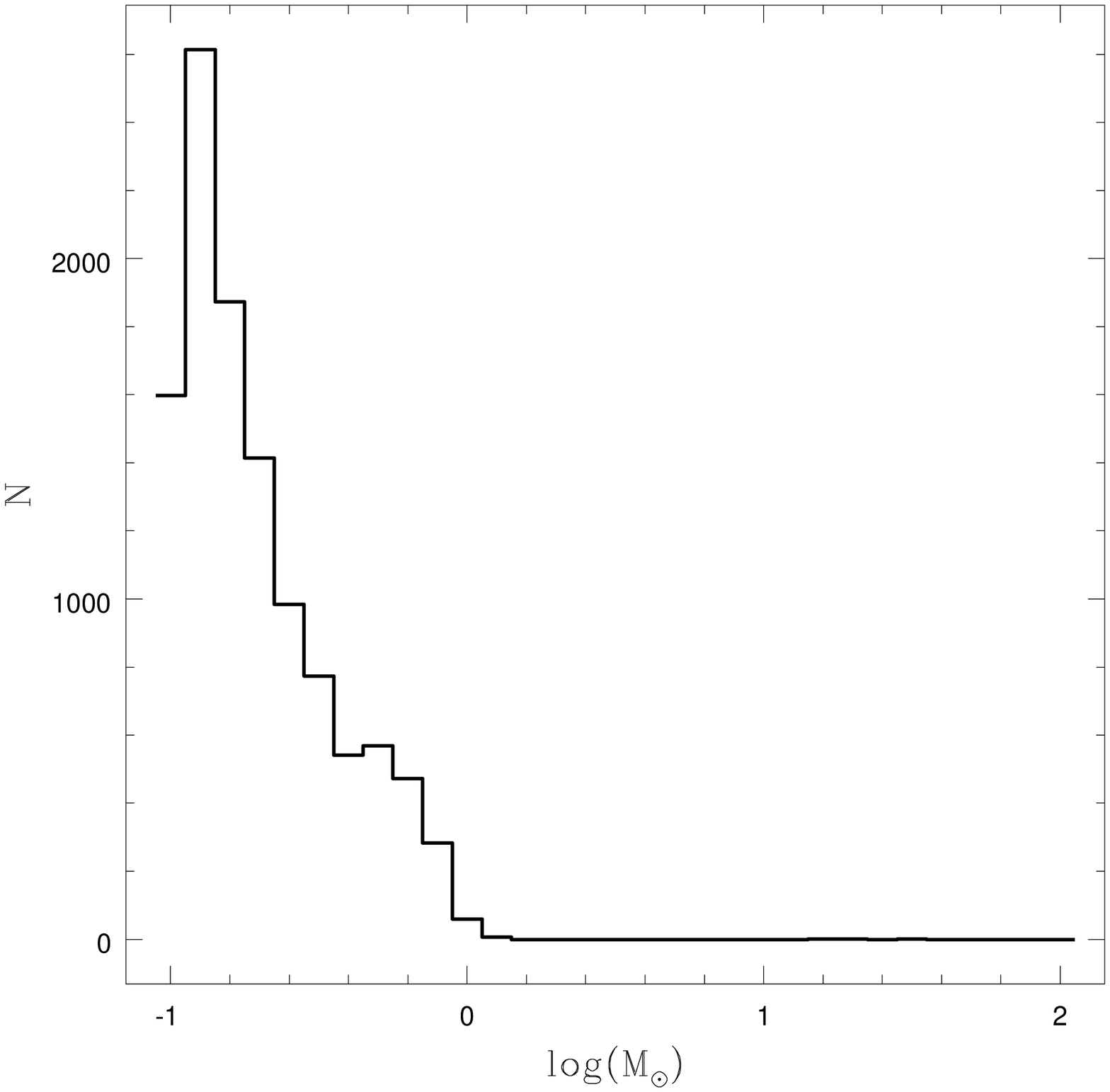}{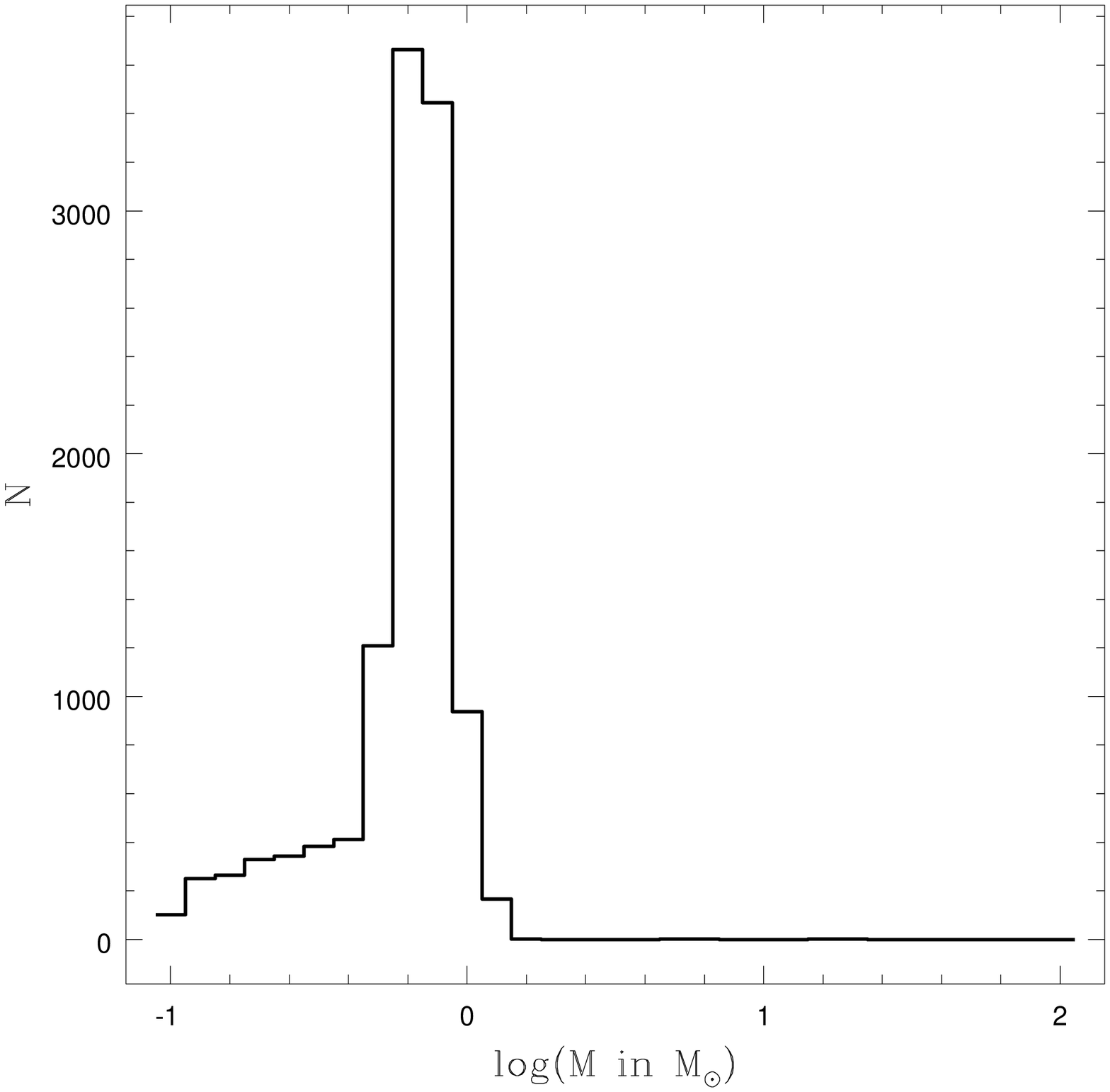}
\epsscale{0.45}
\plotone{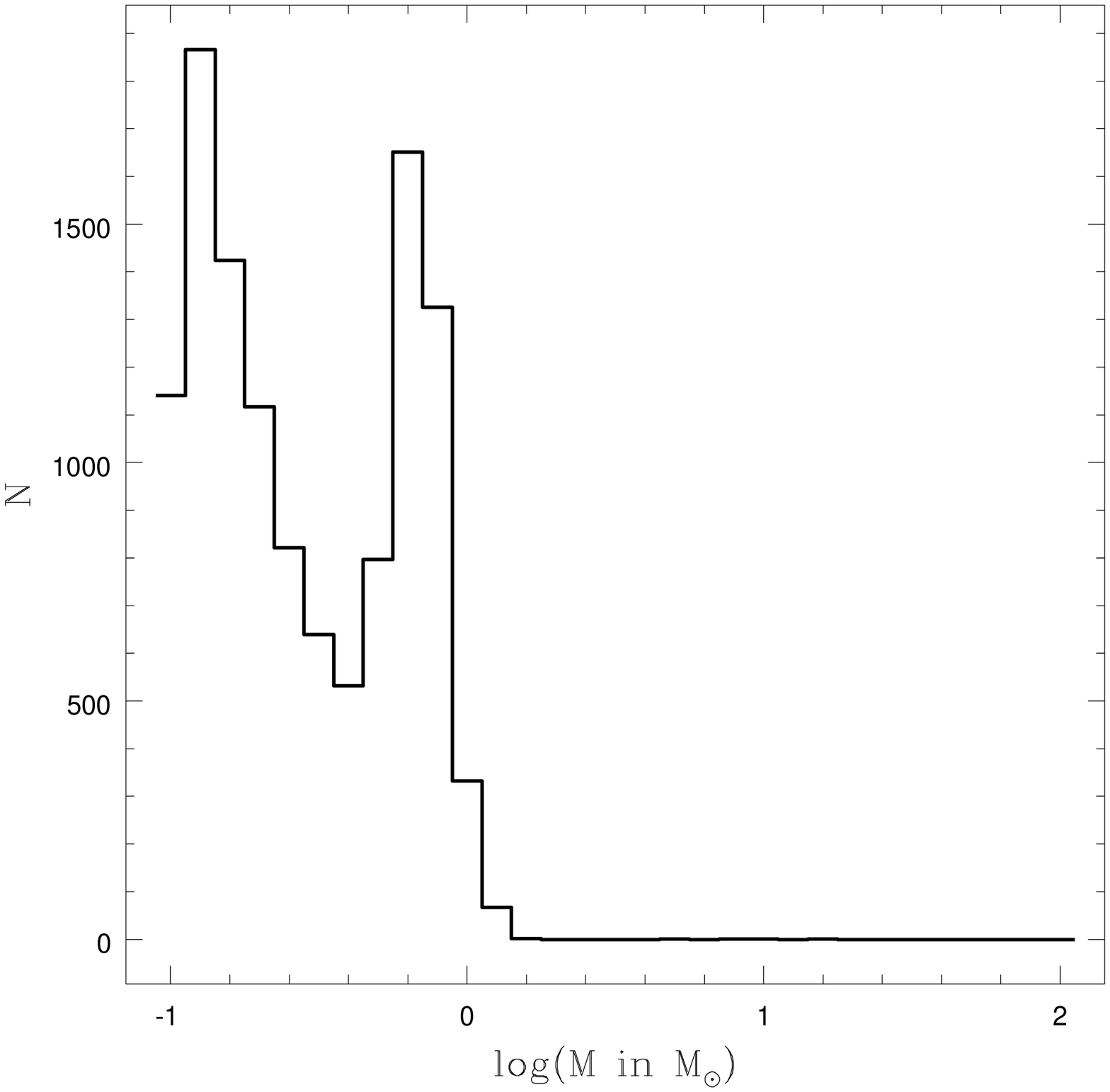}
\epsscale{1.0}
\caption[Final Mass Histograms]{Histograms displaying the final mass
  function for the clusters SP (top left), DC (top right) and Ca (bottom).
  \label{fig:FinMF}}
\end{figure}

Finally, we consider the stellar populations of the clusters.  In
Figure~\ref{fig:FinMF} we present, as histograms, the final mass functions for
SP, DC and Ca.  In the case of SP there has been very little evolution in the
mass function.  The number of $0.7-1 M_{\odot}$ stars has been slightly
enhanced due to the evolution of high-mass stars into white dwarfs but the
stellar population is still dominated by low-mass main sequence stars
(Table~\ref{tab:FinStel}).  By contrast both DC and Ca are very strongly
peaked at $0.7-1 M_{\odot}$.  This peak is composed mainly of C-O white dwarfs
with intermediate-mass progenitors and which make up a much greater fraction
of the population than they do in SP.  Indeed DC is dominated by such compact
remnants (Table~\ref{tab:FinStel}).  The strong double peak in the mass
function of Ca is also to be noted as it resembles the combination of the
final mass function for SP and DC, as would be expected for a cluster formed
from a combination of two such IMFs.  These mass functions are all quite
different and indeed all three types of clusters should be distinguishable by
obtaining accurate white dwarf to main sequence star number ratios.  Recent
observations of NGC 6752 \citep{NGC6752} show an anomalously high mass-to-light
ratio in the core which might be indicative of a large population of
low-luminosity, compact objects such as old white dwarfs.  A early, massive
generation could be one way of producing such an excess although much
more study is needed before we would seriously propose this scenario
as an option in particular cases.

\section{Conclusions}

We have considered the plausibility of a two-generation formation
scenario for galactic globular clusters from the standpoint of
dynamics and to a lesser extent stellar populations.  Based on our
simulations we expect two-generation clusters to be long-lived objects
that at the current age of the galactic globular cluster population
are very dynamically similar to their single-generation counterparts.
We tentatively propose that two-generation clusters can undergo core
collapse at a similar ime to the single generation clusters, although
more investigation is needed on this front.  We expect two-generation
clusters with top-heavy IMF first-generations to have different
stellar populations and specifically we predict that two-generation
clusters with a top-heavy IMF would be strongly enhanced in white
dwarf stars.  Clusters observed to have such an enhancement would be
candidates for the D'Antona and Caloi two-generation formation
scenario.

There are two areas in particular that could use further expansion.  The first
is the treatment of the ejecta in the initial generation.  In particular a
more complete treatment of the gas and a full exploration of different star
formation efficiencies would be useful in order to make perdictions about
young clusters.  In particular it has been shown \citep{GasML2006} that rapid
loss of gas from a young cluster can leave the remaining stellar component out
of equilibrium.  The loss of the non-star-forming fraction of the ejecta from
the first generation could have a similar result on the initial state of a two
generation cluster and deserves exploration.  The second area is to perform
further simulations to pin down exactly when a two-generation cluster will
experience core collapse.

Based on our simple simulations, we find that two-generation clusters are
plausible on the basis of dynamics and could form a fraction of the population
of galactic globular clusters.  Interestingly \cite{Karakas06} show that AGB
stars may not be capable of producing sufficient helium enhancement to produce
the BHB of NGC 2808 without violating other constraints (most importantly
$C+N+O \sim const.$).  Thus, although we find two-generation clusters to be a
dynamical possibility, the application of the two-generation scenario to the
motivating example cluster remains in doubt.

\appendix

\section{Appendix A}

The top-heavy IMF from \cite{DandC2004} is not included
in the standard set of Starlab IMF prescriptions and we have encoded it
ourselves.  First we fix the constants $c_{1}$ and $c_{2}$ by the number of
RHB stars in NGC 2808, the specific choice of $M_{B}$, $\alpha$ and $\beta$
and the criterion $c_{1}M_{B}^{-(1+\alpha)} = c_{2}M_{B}^{-(1+\beta)}$.  This
yields:
\begin{displaymath}
c_{1} = \frac{dN}{dM}\Bigg|_{M_{RHB}} \times M_{RHB}^{-(1+\alpha)} \qquad
\qquad c_{2} = c_{1}M_{B}^{(\beta - \alpha)}
\end{displaymath}
We can then calculate the number of stars in the $\alpha$ and $\beta$ regimes:
\begin{displaymath}
N_{\alpha} = c_{1}\frac{M_{B}^{-\alpha} - M_{l}^{-\alpha}}{-\alpha} \qquad \qquad
N_{\beta} = c_{2}\frac{M_{up}^{-\beta} - M_{B}^{-\beta}}{-\beta}
\end{displaymath}
The mass for an individual node is then generated using the prescription:
\begin{displaymath}
M = \left\{ \begin{array}{ll}
M_{l}[(((\frac{M_{B}}{M_{l}})^{-\alpha} - 1)P_{m} + 1)^{\alpha}] &
P_{mr} \leq \frac{N_{\alpha}}{N}\\
M_{B}[(((\frac{M_{up}}{M_{B}})^{-\beta} - 1)P_{m} + 1)^{\beta}] &
P_{mr} > \frac{N_{\alpha}}{N}
\end{array} \right.
\end{displaymath}
where $P_{mr}$ and $P_{m}$ are random numbers between zero and one which
determine respectively the mass regime a star will be in ($\alpha$ or $\beta$)
and what its mass in that regime will be.  This mass-function is defined
entirely by the choice of $\alpha$, $\beta$ and $M_{B}$ and produces the
needed enhancement in high-mass stars.

\clearpage

\clearpage

\clearpage

\clearpage

\clearpage

\clearpage

\clearpage

\clearpage

\begin{table}[!hbp]
\centering
\footnotesize{
\begin{tabular}[c]{c|c c|c c}
\multicolumn{5}{c}{Simulation Parameters} \\
\hline
{} & \multicolumn{2}{|c|}{$1^{st}$ gen} &
\multicolumn{2}{c}{$2^{nd}$ gen} \\
\hline
Run ID & $N$ & $M_{tot}$($M_{\odot}$) & $N$ &
$M_{tot}$($M_{\odot}$) \\
\hline
SP & 12000 & 4275 & - & - \\
DC & 12000 & 26312 & - & - \\
\hline
Ca & 4096 & 9093 & 8350 & 3132 \\
Cb & 4096 & 9053 & 8400 & 3061 \\
Cc & 4096 & 9032 & 8300 & 2982 \\
\hline
\end{tabular}
}
\centering
\caption[Simulation Parameters]{The parameters used to initialize our
  eight simulations.  ID identifies the run.  The total particle
  number and total mass are then given for the first and second generation
  respectively (the mass of the second generation matches the mass lost in the
  first). \label{tab:InPar}}
\end{table}

\begin{table}[!hbp]
\centering
\footnotesize{
\begin{tabular}[c]{c|c c c c c c c}
\multicolumn{8}{c}{Final Physical Parameters} \\
\hline
ID & $N_{f}$ & $M_{f}$($M_{\odot}$) & $T_{RH}$(Myr) & Final Age(Gyr) &
$M_{10\%in}$($M_{\odot}$) & $M_{10\%fin}$($M_{\odot}$) & $K_{esc}$ (N-body) \\
\hline
SP & 11189 & 2684 & 930.2 & 12.1 & 0.5407 & 0.5147 & 7.376E-2 \\
DC & 11507 & 7308 & 413.9 & 11.3 & 4.537 & 0.8864 & 2.307E-4 \\
\hline
Ca & 11718 & 4336 & 799.8 & 12.2 & 2.053 & 0.7689 & 7.650E-3 \\
Cb & 11842 & 4370 & 736.3 & 12.4 & 2.072 & 0.7693 & 1.764E-3 \\
Cc & 11661 & 4359 & 738.3 & 12.4 & 2.120 & 0.7744 & 1.831E-3 \\
\hline
\end{tabular}
}
\caption[Physical parameters]{Column (1) gives the run ID, (2) is the
  final particle number, (3) is the final mass of the system, (4) is
  the initial half-mass relaxation time, (5) is the final age of the
  cluster, (6) and (7) respectively the initial and final cut-off masses for
  stars to be included in the calculation of the massive Lagrangian radii and
  (8) is the kinetic energy of all stars with a square of velocity greater than
  four times the RMS velocity. \label{tab:FinPar}}
\end{table}

\begin{table}
\centering
\footnotesize{
\begin{tabular}[c]{c|c c c c c}
\multicolumn{6}{c}{Final Stellar Populations} \\
\hline
Type & SP & DC & Ca & Cb & Cc \\
\hline
Main Sequence & 10688 & 3886 & 8736 & 8844 & 8587 \\
White Dwarfs & 461 & 7453 & 2896 & 2903 & 2997 \\
Neutron Stars & 0 & 30 & 15 & 16 & 11 \\
Black Holes & 2 & 4 & 3 & 2 & 1 \\
Other & 38 & 134 & 68 & 77 & 65 \\
\hline
\end{tabular}
}
\caption[Final Stellar Populations]{The final stellar populations for
  each simulation.  ``Other'' includes subgiant, Hertzprung gap,
  horizontal branch etc.  \label{tab:FinStel}}
\end{table}

\end{document}